\documentclass[lettersize,journal]{IEEEtran}
\usepackage{amsmath,amsfonts}

\usepackage{algorithm}
\usepackage{algorithmic}

\usepackage{multirow}
\usepackage{booktabs}

\usepackage{listings}
\usepackage{xcolor}

\lstdefinelanguage{json}{
    basicstyle=\ttfamily\footnotesize,
    stringstyle=\color{red},
    commentstyle=\color{gray},
    showstringspaces=false,
    breaklines=true,
    frame=single,
    morestring=[b]",
    morecomment=[l]{:},
    literate=
     *{0}{{{\color{blue}0}}}{1}
      {1}{{{\color{blue}1}}}{1}
      {2}{{{\color{blue}2}}}{1}
      {3}{{{\color{blue}3}}}{1}
      {4}{{{\color{blue}4}}}{1}
      {5}{{{\color{blue}5}}}{1}
      {6}{{{\color{blue}6}}}{1}
      {7}{{{\color{blue}7}}}{1}
      {8}{{{\color{blue}8}}}{1}
      {9}{{{\color{blue}9}}}{1}
}

\usepackage{algorithmic}
\usepackage{algorithm}
\usepackage{array}
\usepackage[caption=false,font=normalsize,labelfont=sf,textfont=sf]{subfig}
\usepackage{textcomp}
\usepackage{stfloats}
\usepackage{url}
\usepackage{verbatim}
\usepackage{graphicx}
\usepackage{cite}
\usepackage{multirow}

\begin{document}

\title{An O-RAN-Assisted MARL Approach for Dynamic Sidelink and Infrastructure Selection in V2X Communications}

\author{Maria Barbosa, Kelvin Lopes Dias%
\thanks{Copyright © 2026 IEEE. Personal use of this material is permitted. However, permission to use this material for any other purposes must be obtained from the IEEE by sending a request to pubs-permissions@ieee.org.}%
\thanks{Maria Barbosa and Kelvin Lopes Dias are with the Centro de Informática, Universidade Federal de Pernambuco, Recife-PE, Brazil (e-mail: mksb@cin.ufpe.br; kld@cin.ufpe.br).}%
}


\maketitle

\begin{abstract}
Future applications in the 6G-based Internet of Vehicles (IoV) will leverage sidelink (SL) transmissions in Vehicle-to-Everything (V2X) scenarios. However, SL-based direct communication can significantly increase interference among vehicles and between vehicles and other entities of the Intelligent Transportation System (ITS). Thus, both Vehicle-to-Vehicle (V2V) communications and Vulnerable Road Users (VRUs) uplink resources may be degraded or subject to starvation. Existing solutions primarily focus on improving resource allocation and pair selection. Nonetheless, they lack a comprehensive approach to tackle the communication modes and the entire network. To address these challenges, this paper leverages Open RAN to manage V2X communication and proposes a multi-agent reinforcement learning (MARL) resource-aware system. Open RAN provides control loops through a global view of the network and also an open interface-based framework for machine learning models applied to resource decision-making. Meanwhile, the MARL model aims to mitigate interference, optimize resource usage, and enhance quality of service by optimally selecting between sidelink and network (V2N) transmissions. To reduce system complexity, this work employs a clustering strategy. Each agent manages a group of pairs, rather than assigning one agent to each pair. The solution supports this design by adopting a centralized training with decentralized execution (CTDE) approach, empowered by Open RAN. The O-RAN Alliance specifications require pre-trained models. To comply, the strategy uses offline training and an off-policy approach, where each agent stores experiences for fine-tuning. For benchmarking, the baselines include a single-agent approach and two heuristics: a Signal-to-Interference-plus-Noise Ratio (SINR)-based scheme and the Best Channel Quality Indicator (CQI) policy. The MARL approach reduces average loss by 21\% and latency by 19\% in vehicle-only scenarios, and by 18\% and 30\% in VRU coexistence scenarios, respectively.

\end{abstract}


\section{\textcolor{black}{Introduction}}
\label{sec:intro}

\IEEEPARstart{I}{ntelligent} Transportation Systems (ITS) aim to enhance road safety and operational efficiency, driving the evolution of the connected vehicles paradigm and, consequently, the emergence of the Internet of Vehicles (IoV). In this context, vehicle-to-everything (V2X) communication plays a crucial role in meeting the stringent requirements of low latency, high reliability, and high data throughput \cite{v2x_intro}. V2X enables the deployment of emerging vehicular applications, such as remote, advanced, and cooperative driving, road warning, platooning, and see-through systems \cite{v2x_cases}.

To enhance V2X communication, the 3rd Generation Partnership Project (3GPP) proposed Sidelink (SL) communication, which enables vehicles to communicate directly with each other. The 3GPP  introduced the SL concept in LTE Release 12 and added support for V2X communication in Release 14. Nevertheless, only the 5G NR Release 16 expanded the support to include the aforementioned advanced use cases \cite{3gpp_sidelink}. SL uses part of the uplink radio resources \cite{sidelink_use_ul}. In this context, resource sharing among Vulnerable Road Users (VRUs), which in ITS refers to non-motorized road users such as pedestrians and cyclists, as well as vehicle-to-infrastructure (V2I), vehicle-to-vehicle (V2V), and vehicle-to-network (V2N) communications, leads to mutual interference and reduces overall system performance. Furthermore, the limited availability of uplink resources also constrains the system \cite{limited_resources}.

To manage resource allocation (RA), 3GPP defines two control modes for SL operation. In Mode 1, the gNB handles RA directly. However, this centralized approach suffers from limited visibility into link conditions and the specific traffic demands of each V2X communication. In Mode 2, the UE autonomously allocates SL resources, offering flexibility, but it struggles to resolve resource conflicts effectively, resulting in increased interference and degraded communication performance \cite{Modes_SL}.

The literature has addressed these issues by proposing models for path estimation, such as the one in \cite{abbas_road-aware_2019}, which aims to optimize resource utilization by enabling vehicles to list their neighbors and next hops in the Infrastructure-Based Hybrid Road-Aware Routing Protocol (IARAR). They also collect vehicle speed to estimate path duration and potential communication time over the link. However, despite the flexible infrastructure, the authors neither consider the current network state, such as resource availability or interference levels, nor propose any balancing scheme for link utilization. In contrast, \cite{ashfaq_uplink_2024} presents the Uplink Resource Shared Interference Mitigation Scheme (URSIMS), which consists of two heuristic algorithms: one to handle resource sharing among different zones and another to define signal power control. The algorithm uses distance to calculate path loss and, together with fading, determines the best Signal-to-Interference-plus-Noise Ratio (SINR) between D2D pairs and a regular UE. Thus, this problem can be approached as a minimization issue. Nevertheless, the solution does not benefit from either the flexibility or the radio information measurements.

Recently, the Open Radio Access Network (Open RAN), introduced by the O-RAN Alliance in 2018 with the aim of reducing vendor lock-in and fostering innovation within the RAN ecosystem, has also been adopted as a promising architecture to address key challenges in V2X communications \cite{oran_use_cases}. This paradigm promotes multi-vendor interoperability and programmability in mobile networks through a modular architecture based on open interfaces. The architecture defines two categories of RAN Intelligent Controller (RIC): Non-Real-Time (Non-RT RIC) and Near-Real-Time (Near-RT RIC). The Non-RT RIC operates with control times exceeding one second, is part of the Service Management and Orchestration (SMO) framework, and hosts microservices, referred to as rApps. The Non-Real-Time is responsible for policy generation and the lifecycle management of Artificial Intelligence/Machine Learning (AI/ML) models. The Near-RT RIC acts in control times between 10 and 1000 ms and manages the interfaces with the E2 nodes (e.g., 5G gNodeBs), enabling microservices, called xApps, to monitor and control functionalities, such as handover management \cite{ho_icoin}, load balancing \cite{oran_load_rl_2} and traffic steering \cite{policy_ts}, thereby facilitating AI-driven decision-making and dynamic policy enforcement.

In \cite{linsalata_addressing_2024}, the proposal leverages the orchestration capabilities of O-RAN elements using data exchange in the Below 6 GHz (sub-6 GHz) band to facilitate the reliable selection of relay nodes operating in mmWave. To illustrate the advantages of the solution, an xApp was designed for relay selection, incorporating Line of Sight (LoS) recognition and distance measurement. The authors in \cite{hammami_multi-agent_2022} aim to address the scarcity of shared resources between the V2V and V2I problems, highlighting the risk of interference when both entities reuse the same resources. To tackle this, the proposed solution uses a multi-agent reinforcement learning (MARL) model aiming to select the most suitable resource block groups to minimize interference and increase transmission rates.

Nevertheless, the adopted approach in \cite{hammami_multi-agent_2022} assigns one agent per link, which becomes unfeasible in an IoV network since the number of vehicles can grow indefinitely, leading to processing overload. While \cite{gjeci_towards_2024} proposes an Open RAN-based solution for radio resource management, considering connected and autonomous vehicles (CAVs) as E2 endpoints. The performance evaluation compared Mode 2 with the proposed solution. However, the authors do not indicate what happens if resources are exhausted. 

\textcolor{black}{Existing solutions in the literature typically do not jointly address interference mitigation and resource allocation optimization in IoV networks, especially on sidelink communications, nor do they ensure minimum QoS requirements. In addition, the coexistence between vehicles and VRUs, competing for limited resources, is often neglected. Given this, this work aims to jointly mitigate interference and optimize resource allocation under dynamic, resource-constrained conditions. In this context, communication mode selection is used to assign the most suitable mode to each vehicle pair, distributing users across V2V and V2N. Since resources are limited, it is crucial to prioritize their allocation to pairs where sidelink usage is more efficient, while avoiding resource starvation for non-prioritized VRUs and for vehicles operating in dense areas or under conditions of interference and resource scarcity. } 

\textcolor{black}{To address these challenges, it is important to recognize that traditional optimization approaches may struggle to capture the dynamic and large-scale nature of this problem, especially in high-mobility, dense settings and heterogeneous traffic conditions. Moreover, single-agent learning strategies suffer from generalization limitations, while fully decentralized approaches with one agent per link become impractical as network density increases. Therefore, this work models the problem as a MARL task and introduces a clustering strategy in which each agent is responsible for a group of vehicles, improving scalability while preserving coordination among decisions.}

\textcolor{black}{To enable this solution, this work also adopts the Open RAN architecture, which, unlike traditional controllers, explicitly defines the lifecycle management of AI/ML models and specifies different deployment scenarios that distribute training and inference processes between the Non-RT RIC and the Near-RT RIC.}

\textcolor{black}{The proposed solution adopted deployment scenario 1.2 among those defined by O-RAN, as it allows structuring the learning process according to the Centralized Training with Decentralized Execution (CTDE) paradigm. In this scenario, the Non-RT RIC serves as the training host and manages continuous operations, which are critical and computationally intensive, however, since it operates at timescales greater than 1s, it avoids impacting control latencies. While the Near-RT RIC is responsible for inference and operates with lower latency, it is critical for the management and control of IoV networks. Finally, another challenging factor is that the O-RAN architecture requires pre-trained models for inference, therefore, the present proposal introduces a methodology for offline MARL training.}

\textcolor{black}{Additionally, an essential aspect of the Open RAN architecture is the modularization of control and monitoring into separate xApps, enabling a modular solution that meets control latency requirements. In this way, the present proposal leverages this characteristic by functionally separating the monitoring, inference, and control components. This decomposition allows reduces the coupling between data collection, inference, and action execution.}

\textcolor{black}{Furthermore, the architecture enables, through standardized signaling and interfaces, especially the E2 interface, the implementation of control loops, decision-making, and continuous monitoring. This monitoring provides a structured, standardized view of the network state by collecting and storing user measurements, which are directly used to construct the state space of the MARL model. These data also enable continuous model refinement through fine-tuning based on accumulated operational experience. In this context, reinforcement learning has shown promise in Open RAN environments by enabling continuous policy adaptation based on real operational data. This adaptability is crucial in vehicular networks, where high mobility increases network dynamism.} In short, the main contributions of the paper are:

\begin{itemize}

\item To address interference and limited sidelink availability, the problem is modeled as a Markov Decision Process (MDP). Therefore, to optimize communication mode selection and resource allocation, ensuring minimum QoS requirements, we use a MARL-based strategy. This approach enables dynamic assignment of V2N and V2V links according to the suitability of each communication pair. Additionally, the proposal evaluates system performance in two scenarios: a vehicle-only network and one with coexisting VRUs competing for resources. Results show that the approach achieves a more efficient balance between uplink and downlink usage in both cases. Specifically, in the vehicle-only scenario, approximately 80\% of communication pairs achieved SINR values around 28 dB, outperforming the single-agent strategy.

\item Since the Open RAN architecture enables lifecycle management of ML models, supports the CTDE strategy, and, given that the O-RAN specification does not permit untrained models in production, the methodology includes an initial offline training phase using a conservative approach, followed by conservative fine-tuning updates that are better suited to offline environments.

\item To avoid over-provisioning intelligent control loops in the Open RAN architecture, the proposal groups users by assigning one per agent. During training, each agent stores its experiences in a retraining buffer, enabling cooperative learning. The solution leverages the QMIX strategy, which enables agents to share their experiences through a Mixer network during training/fine-tuning. The evaluation reveals that increasing the number of agents results in higher computational resource consumption, with a 14.94\% increase in memory and a 9.94\% decrease in CPU usage when comparing 50 agents to 75 agents. In the vehicle-only scenario, adding more agents did not necessarily improve QoS metrics, as evidenced by the 8.6\% increase in mean latency from 50 to 75 agents. However, in the more complex and heterogeneous scenario involving both vehicles and VRUs, the configuration with the highest number of agents achieved an 8\% reduction in mean latency compared with 50 agents, highlighting a clear trade-off between QoS and computational resource usage.
\end{itemize}

The remainder of this paper follows this structure. Section \ref{sub_sec_sidelink} introduces the overview concepts about SL. In Section \ref{sys_model_prob_form}, we model the system and formulate the problem. Section \ref{sec:prop} details the proposed solution and its architectural components. Section \ref{sec:results} presents the evaluation scenarios and results. Finally, Section \ref{sec:conc} concludes the paper and outlines future research directions.

\section{An overview of sidelink communications}
\label{sub_sec_sidelink}

In V2X systems, the V2N communication mode provides vehicular communication to the network services through the Uu interface, while the direct V2V transmissions utilize the PC5 interface. Introduced in 3GPP Rel-12 to support Proximity Services (ProSe) services, PC5 enabled early D2D communication and is now extensively employed in V2X scenarios. It is well-suited for high-mobility environments and offers stable performance even under high node density. Additionally, PC5 ensures low latency and high reliability, though it has a shorter transmission range than Uu. Unlike V2N, V2V enables out-of-coverage communication, which enhances road safety by allowing data exchange independent of gNodeB connectivity.

The allocation of Sidelink resources involves managing the Physical Sidelink Control Channel (PSCCH), Physical Sidelink Shared Channel (PSSCH), and Physical Sidelink Feedback Channel (PSFCH), which are responsible for resource scheduling, D2D data transmission, and ACK/NACK feedback signaling, respectively. These resources may be centrally managed by the gNodeB (Mode 1) or autonomously by the UE (Mode 2), with both modes (pre-)configured by the Radio Resource Control (RRC) layer. Sidelink uses a shared bandwidth for transmission and reception, structured into time-frequency resource pools \cite{v2x_cases} \cite{Modes_SL}. 

\begin{figure}[ht!]
    \centering
    \includegraphics[width=1\linewidth]{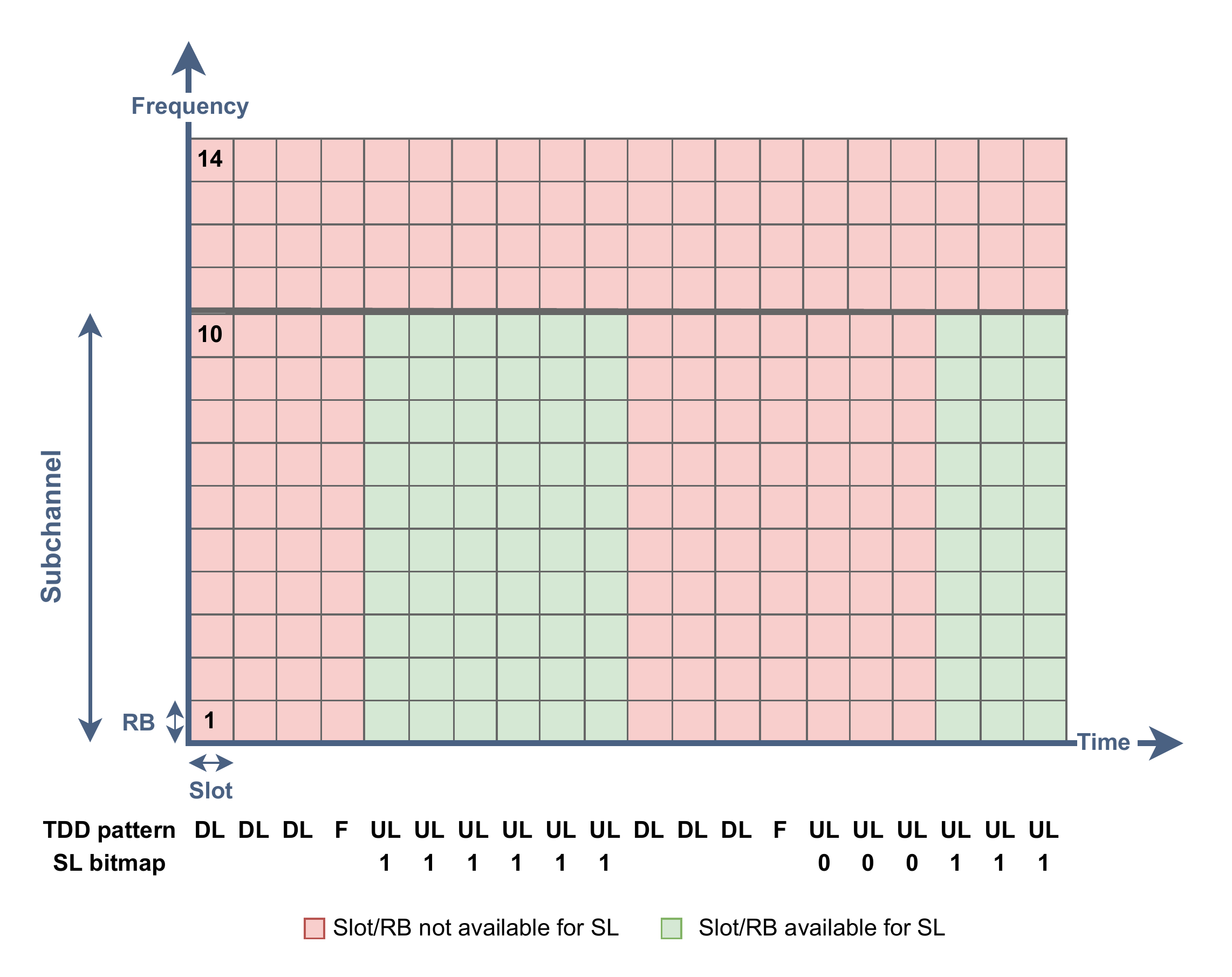}
    \caption{Frame structure and sidelink bitmap.}
    \label{fig:sl_bitmap}
\end{figure}

In the time domain, the Sidelink bitmap (pre-)configured by the RRC periodically defines the available slots. This bitmap is combined with the TDD pattern in TDD systems to determine usable slots \cite{v2x_cases}. For example, a subcarrier spacing of 60 kHz with a 10 MHz bandwidth and a TDD pattern of [DDDFUUUUUU] and a bitmap of [111111000111] define the sidelink slots, as shown in Figure \ref{fig:sl_bitmap}. The figure shows that SL operates using uplink (UL) resources, but only the available ones. The system groups the resources into contiguous subchannels in the frequency domain, each containing $N$ resource blocks (RBs). For NR V2X, $N$ may assume 10, 15, 20, 25, 50, 75, or 100 RBs. According to \cite{sidelink_use_ul}, the last four RBs are reserved and unavailable for Sidelink operation. 

\section{System Model and Problem Formulation}
\label{sys_model_prob_form}

\subsection{System Model}

Consider a V2X scenario of $V$ vehicles, represented by the set $V = \{1, 2, \dots, v, \dots, V\}$, moving along a highway supported by a base station (BS).

\textbf{\textit{Definition 1 – Correlation between available and used radio resources}}: 
The BS has a total of $F$ Physical Resource Blocks (PRBs) available in time, where:
\begin{equation}
    F = F_{\text{UL}} + F_{\text{DL}}, \quad \text{with } F_{\text{DL}} > F_{\text{UL}}.
\end{equation}

The sidelink communication (PC5) consumes a portion of the uplink resources:
\begin{equation}
    F_{\text{UL}} = F_{\text{UL}}^{\text{BS}} + F_{\text{SL}},
\end{equation}
where $F_{\text{UL}}^{\text{BS}}$ corresponds to uplink resources used for conventional V2N communication, and $F_{\text{SL}}$ is the portion allocated for sidelink (V2V).

Each V2V pair $(i,j) \in P_{\text{V2V}}$ requires $f_{i,j}$ PRBs to maintain stable communication using the PC5 interface. Therefore, the total number of PRBs consumed by all active V2V pairs is:
\begin{equation}
    \Gamma_{\text{V2V}} = \sum_{(i,j) \in P_{\text{V2V}}} f_{i,j},
\end{equation}
and all this demand is drawn from $F_{\text{SL}} \subset F_{\text{UL}}$.

On the other hand, V2N pairs $(i,j) \in P_{\text{V2N}}$ consume resources from both uplink and downlink:
\begin{equation}
    \Gamma_{\text{V2N}}^{\text{UL}} = \sum_{(i,j) \in P_{\text{V2N}}} f_{i,j}^{\text{UL}}, \quad 
    \Gamma_{\text{V2N}}^{\text{DL}} = \sum_{(i,j) \in P_{\text{V2N}}} f_{i,j}^{\text{DL}},
\end{equation}
with total V2N demand:
\begin{equation}
\label{eq:reso_pairs_v2n}
    \Gamma_{\text{V2N}} = \Gamma_{\text{V2N}}^{\text{UL}} + \Gamma_{\text{V2N}}^{\text{DL}}.
\end{equation}

 $\Gamma_{\text{V2N}}^{\text{UL, pairs}}$ and $\Gamma_{\text{V2N}}^{\text{DL, pairs}}$ refer to uplink and downlink resources consumed by vehicle pairs using the Uu (V2N) interface ($P_{\text{V2N}}$).

If the sidelink demand exceeds the available portion, i.e., $\Gamma_{\text{V2V}} > F_{\text{SL}}$, the network experiences overload in the sidelink domain, which may result in packet losses and increased latency. This study assumes this condition holds and investigates mechanisms for dynamic balancing between the PC5 (V2V) and Uu (V2N) interfaces.

\textbf{\textit{ Definition 2 – V2V Interference}}: Each V2V pair \((i,j)\) experiences a Signal to Interference plus Noise Ratio (SINR) given by:

\begin{equation}
    \text{SINR}_{i,j} \approx \frac{PW_{i,j} h_{i,j}}{I_{\text{total}} + N_0}, 
    \label{eq:sinr}
\end{equation}

where $PW_{i,j}$ is the transmission power from vehicle $i$ to vehicle $j$, \( h_{i,j}\) is the channel gain between them. The denominator includes the noise power $N_0$ and the total interference from all other pairs $(m,n)$:

\begin{equation}
    I_{\text{total}} = \sum_{(m,n) \neq (i,j)} PW_{m,n} h_{m,n}
\end{equation}

As the number of vehicles $V$ increases, more V2V pairs share the same limited PRBs, raising the interference level.

Furthermore, using the Uu interface in V2X communications becomes more challenging when users move to the edge of the in-coverage area. In such cases, the total received power ($PW_{r}$) is:

\begin{equation}
        PW_{r} = PW_{t} - PL,
\end{equation}

where $ P_{t}$ is the transmitted power, and $PL$ is the path loss. Under line-of-sight conditions \cite{path_loss}, the path loss is given by:

    \begin{equation}
        PL= 22log(d_{3D}) + 20log(f_{c}) + 28.0
    \label{FSPL}
    \end{equation}

Thus, signal degradation naturally occurs as the distance between the vehicle $v$ and the gNodeB increases.
 
\textbf{\textit{Definition 3 – Coexistence with Vulnerable Road Users (VRUs)}}: 
This work assumes that VRUs run applications that require significantly more downlink than uplink resources, such as video/music streaming or real-time alerts. Therefore, their uplink demand is considered negligible.

As a result, vehicles experience resource competition not only with other V2V pairs but also with VRUs for the V2N communication resources. Specifically, the total consumption of PRBs in the V2N domain becomes:

\begin{equation}
    \Gamma_{\text{V2N}} = \Gamma_{\text{V2N}}^{\text{UL, pairs}} + \Gamma_{\text{V2N}}^{\text{DL, pairs}} + \Gamma_{\text{V2N}}^{\text{DL, VRUs}},
\end{equation}

where $\Gamma_{\text{V2N}}^{\text{DL, VRUs}}$ corresponds to downlink resources consumed by VRUs.

Moreover, VRUs increase the overall interference level in the downlink, as expressed in Equation~\ref{eq:sinr}, by contributing to the total interference term $I_{\text{total}}$.

\subsection{Problem Formulation}

We aim to minimize the resource consumption problem, particularly the uplink PRBs usage, while maximizing the SINR, ensuring the loss and latency requirements by optimizing the communication mode selection by assigning the most suitable modes for each pair. To this end, we formulate the problem as a Markov Decision Process (MDP) for MARL applied to balancing in V2V and V2N communication pairs. The MDP tuple (S, A, R, p, $\gamma$) represents the environment, where S is the state space, A is the action space, R is the rewards, p is the transition probabilities, and $\gamma$ is the discount factor.

Most existing solutions rely on single-agent strategies, where a centralized model performs predictions and makes decisions. However, the Open RAN architecture enables centralized control over the entire network. Due to the dynamic nature of IoVs, single-agent strategies often face generalization challenges. In these strategies, an agent selects an action $A_t$ for a given state $S_t$ and receives a reward $R_t$. If $v$ vehicles share similar states, the agent tends to select the same action $A_t$ for each, aiming to maximize its reward. This behavior can lead to overloading one link while underutilizing others. Multi-agent architectures provide a more suitable solution by distributing tasks among agents and supporting collaborative learning.

This study addresses balancing challenges using MARL within the O-RAN framework in V2X scenarios. The problem involves dynamic network topologies and stringent latency and reliability requirements in vehicular networks. High vehicle mobility on highways further complicates the management of the PC5 interface. Additionally, the high density of vehicles in V2X scenarios exacerbates scalability challenges.

This work adopts a MARL approach to meet the minimum QoS requirements for vehicle applications. Markovian principles provide a suitable structure for modeling MARL in dynamic vehicular networks \cite{hammami_multi-agent_2022, oran_ml_xapp}. The proposed agents implement the QMix algorithm \cite{qmix_base} with a conservative strategy \cite{cql}. 

\textcolor{black}{Let $P$ be the set of communication pairs $(i,j)$ existing in the network, with $|P|>1$. Solutions that use one agent per pair tend to grow indefinitely, such that $N \rightarrow |P|$ leads to scalability and exploration issues and complicates agent coordination and communication \cite{marl_scale}. The clustering strategy subdivides $P$ into $N$ subsets, where $N$ is the number of agents in the network. Each agent $k$ manages a group $G_{pk}$, that represents the subset of pairs assigned to agent $k$, formed using a round-robin strategy that balances the number of pairs among agents. Group assignments remain fixed while a pair stays active in the network. Nevertheless, the composition of each group naturally evolves over time as pairs form and dissolve.}

The proposed solution limits the number of agents to minimize this scalability problem. So, the study defines three scenarios with 25, 50, and 75 agents for evaluation. This study empirically selects these values to analyze how scalability affects coordination and system performance. This setup enables the observation of how different agent distribution scales affect collaborative learning efficiency, even without a predefined optimal criterion.

Several factors influence load-balancing decisions. From the user perspective, key metrics include path loss (PL), channel quality indicators (CQI), reference signal received power (RSRP), signal-to-interference-plus-noise ratio (SINR), packet loss, and latency. From the base station perspective, resource availability plays a crucial role. The model optimizes user distribution across available resources while minimizing latency and packet loss.

The system considers seven key decision variables defined in the previous subsection. According to Definition 1, the variable $\{F_{SL}\}$ represents resource availability for the sidelink. Definition 2 includes interference and user distance-related factors $\{SINR, RSRP, CQI, loss, latency\}$. Definition 3 introduces downlink resource availability $\{F_{DL}\}$.

The MARL solution defines the state space, action space, and reward function as follows:

\textbf{\textit{(a) State Space ($S$):}} Each agent operates in a zone $A$ for the group $G_{pk}$ and observes only a partial view of the global environment state, associated with $G_{pk}$. Each group includes $n$ links, and the agent state is defined individually per link. Formally, we define the state at time $t$ as:

\begin{equation}
    S_{pk, n}^{t} = \{F_{SL}, S_{i,j}, R_{i,BS}, C_{i,j}, L_i, D_{i,j},F_{DL}\}
\end{equation}




\textbf{\textit{(b) Action Space ($A$):}} \textcolor{black}{The action space is binary, since the proposed MARL model only determines the communication mode assigned to each link. At each decision step $t$, agent $k$, responsible for group $G_{pk}$, independently selects one action for every link $n$ associated with a pair $(i,j) \in G_{pk}$. Each action determines whether the corresponding link operates through V2N or V2V, as defined by:}
\[
A_{k,n}^{t} =
\begin{cases}
0, & \text{use infrastructure (V2N)}\\
1, & \text{use sidelink (V2V)}
\end{cases}
\]
\textbf{\textit{(c) Reward ($R$):}} \textcolor{black}{The reward computation is performed in two stages. First, after each state transition $(S_t,A_t,S_{t+1})$, agent $k$ computes an individual reward $r_{D_{i,j}}^{t}$ for each pair $(i,j) \in G_{pk}$, based on the delay $D_{i,j}$ experienced by that pair. These individual rewards are stored and later aggregated during the offline centralized training phase to compute the global reward, 
\[
R_t = \sum_{(i,j) \in P} r_{D_{i,j}}^{t},
\]
which is used by the QMIX mixing network to optimize the joint action-value function.}

\textcolor{black}{To compute $r_{D_{i,j}}^{t}$, only the communication delay is considered. Preliminary experiments showed a strong correlation between delay and packet loss. Therefore, to avoid optimizing partially redundant metrics, the reward formulation considers only delay, while indirectly improving packet delivery performance. The resulting reward function is defined as:}

\[
r_{D_{i,j}}^{t} =
\begin{cases} 
D_{i,j}^{t} > 30 \text{ ms}, & -0.1 \cdot D_{i,j}^2 - 10 \cdot D_{i,j} + 300 \\[6pt]
D_{i,j}^{t} \leq 30 \text{ ms}, & 10 \cdot \tanh(0.45 \cdot D_{i,j} + 13.6)
\end{cases}
\]

The model rewards the agent positively when it selects appropriate actions, with the equations tuned to maintain reward values around 10. In contrast, the equations for negative rewards apply steeper penalties. When an agent selects an inappropriate action, the model imposes a more substantial penalty, emphasizing the severity of incorrect decisions.

\section{Proposed Multi-Agent-based O-RAN Framework for Connectivity Mode  Selection}
\label{sec:prop}
\begin{figure*}[h!]
    \centering
    \includegraphics[width=0.65\linewidth]{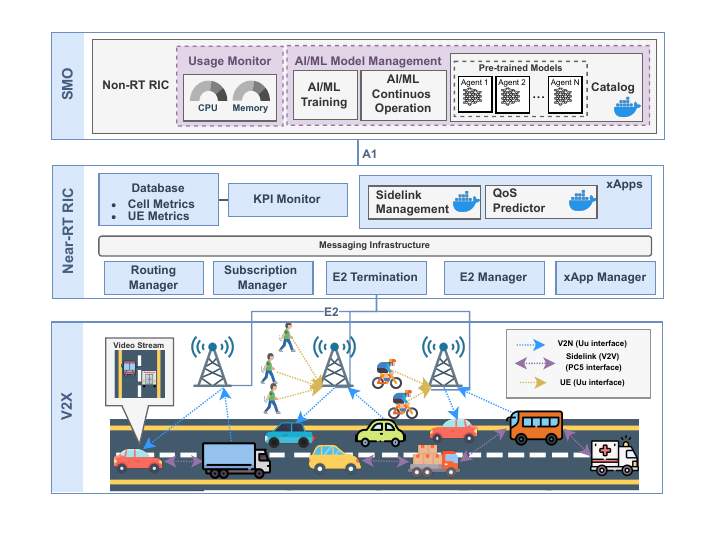}
    \caption{Overview of the proposed solution.}
    \label{fig:proposal}
\end{figure*}

This section presents the proposed solution and its architectural components. As shown in Figure \ref{fig:proposal}, the architecture includes the AI/ML model manager and the computational resource monitor in the Non-RT RIC. \textcolor{black}{Meanwhile, in the Near-RT RIC, the components are modularized into three services: (i) monitoring, through the KPM Monitor xApp; (ii) inference, via the QoS Predictor; and (iii) decision-making, through the Sidelink Manager.}

\textcolor{black}{These xApps interact with the RAN through the standardized E2 interface, which enables control and monitoring loops based on the functionalities exposed by the E2 nodes. In this context, the network state is constructed from metrics reported by the E2 nodes (e.g., SINR, RSRP, CQI), as made available by the E2 interface. Initially, the KPM Monitor subscribes to the desired metrics from the E2 node. The measurements reported by the vehicles are aggregated by the gNodeB, forwarded to the xApp via indication messages, and then stored for use in the inference and decision-making process.}

\subsection{AI/ML Model Manager}

In a multi-agent approach, there could be situations in which an individual agent cannot select an appropriate action. QMix \cite{qmix_base} addresses this issue by leveraging a neural network architecture that models both the agents' individual and global value functions. Each agent has its Q-network, which estimates the value of actions based on local observations. Then, the Mixer network aggregates the individual estimations, taking the global state of the environment as input and producing a joint action value. The Mixer network enforces monotonicity concerning individual Q-values, ensuring that the joint action maximizing the global value maximizes each local action of an agent. This design enables CTDE, making QMix particularly well-suited for collaborative environments such as Open RAN. The training of QMix involves minimizing the following loss function, which measures the mean squared error between the estimated total Q-value and the target value computed from the immediate reward and the Q-value of the next state:

\begin{equation}
	 \mathcal{L}{qmix} = \frac{1}{B} \sum_{i=1}^{B} \left( y^{\text{tot}}i - Q{\text{tot}}(\boldsymbol{\tau}_i, \mathbf{u}_i, s_i; \theta) \right)^2
 \end{equation}

This approach, based on the CTDE paradigm, enables centralized training of the joint value function while preserving decentralized execution among the agents.

Furthermore, no model can be deployed in the Open RAN architecture without prior offline training \cite{ai_workflow_tr}. Therefore, the training and model update phases also incorporated a conservative MARL approach. 

This strategy penalizes action values that fall outside the support of the offline dataset, helping avoid overestimation. The conservative loss function for the multi-agent scenario, 

\begin{equation}
    \mathcal{L}_{\text{cql}} = \frac{\alpha}{N} \sum_{i=1}^{N} \left( \mathbb{E}_{a \sim \pi_i} \left[ Q_i(s, a) \right] - \mathbb{E}_{a \sim D_i} \left[ Q_i(s, a) \right] \right)
\end{equation}
 
Letting $N$ denote the number of agents, $Qi(s,a)$ represents the value function for agent $i$, $\pi_i$ is its current policy, and $D_i$ corresponds to the offline dataset used to estimate action values, and $\alpha$ is the regularization strength. The conservative penalty term is normalized by $N$, allowing the total loss to scale proportionally with the number of agents while maintaining optimization stability.
So, the final loss function of the Conservative QMix (CQMix) is:

\begin{equation}
 \mathcal{L}{\text{cqmix}} = \mathcal{L}{\text{qmix}} + \mathcal{L}_{\text{cql}}
 \end{equation}

During action selection, to balance the conservative training approach with exploration, an $\varepsilon$-greedy policy is adopted:
\begin{equation}
 	\pi(a \mid s) = (1 - \varepsilon) \cdot \mathbb{I}\left[a = \arg\max_{a'} Q(s, a')\right] + \varepsilon \cdot \frac{1}{|\mathcal{A}|}
 \end{equation}
Here, $\varepsilon$ is the exploration rate, $\mathbb{I}[\cdot]$ is the indicator function, and $\mathcal{A}$ denotes the set of available actions.
In this policy, agents select the action with the highest Q-value with probability $1 - \varepsilon$ and choose a random action with probability $\varepsilon$, enabling controlled exploration.
This configuration offers a safe yet effective policy: it remains predominantly conservative (e.g., 95\% greedy actions), while the exploration component helps avoid local optima and compensates for inaccuracies in the Q-value estimates. Moreover, CQL acts as a safeguard against actions outside the dataset distribution. Hence, this approach ensures training safety and stability in offline settings without sacrificing limited exploratory capabilities.
The initial model was offline-trained using data collected during simulations involving only V2N and V2V links, separately, resulting in an initial policy $\pi_0$. However, due to the limited exploration, the resulting model may suffer from poor generalization. Therefore, the model is updated every $n$ executions via fine-tuning, enabling the policy $\pi_n$ to improve based on newly stored experiences in the dataset derived from prior policies $\pi_{k-1}$ \cite{off_to_on_train}. Finally, pre-trained models are stored in a catalog and are made available for offloading to initiate the inference phase.

\subsection{QoS Predictor}

The QoS Predictor (QP xApp) is an xApp responsible for performing ML model inference. To do this, it queries the model catalog to select the most suitable model for the requested prediction. In this work, we evaluate the number of agents in operation at 25, 50, and 75. Thus, the QP xApp sends a request to the AI/ML Manager specifying the desired number of agents and the prediction type (e.g., peer communication mode).

The AI/ML Manager checks for a compatible model and, if available, sends it to the QP xApp along with the execution prerequisites, including the required libraries and their versions. And the formatting and features of the model input state and output vectors. Appendix A presents an example of the model description received by the QP xApp.

\textcolor{black}{During the inference phase, agents operate independently, relying solely on observations of the pairs for which they are responsible, thereby eliminating the need for communication between agents and real-time synchronization. The agents are deployed in the near-RT RIC as part of the QP xApp and are continuously updated during network operation.} 

To initiate the prediction process, the QP xApp waits for a request from the Sidelink Manager. Upon receiving the request, the service identifies the vehicle pairs to evaluate and retrieves the respective state vectors for each pair. The QP xApp then pre-processes the data to fit the ML model input format, executes the inference, and returns the recommended actions to the requester. 

\textcolor{black}{The QP xApp triggers a model update request every 100 processed inference requests per agent. To reduce synchronization costs, the pairs in the network are evenly distributed among agents so that all agents reach the required number of inferences to trigger the update, approximately, at the same time. When all agents reach this threshold, the update request is triggered.}

This request initiates the fine-tuning procedure, in which the agents share their experiences. \textcolor{black}{This procedure causes a synchronization cost linearly proportional to the number of agents, since the collected experiences (e.g. states, actions and rewards) must be aggregated to compute the joint action-value function. However, this step occurs outside the critical real-time cycle. As a result, the proposed approach ensures low-latency inference while maintaining coordinated learning through centralized training.}

Subsequently, the QP xApp performs periodic checks to verify the availability of the updated model. Once the model is ready for offloading, the QP xApp performs the replacement and restarts the inference cycle.

\subsection{Sidelink Management Mechanism}

The sidelink management mechanism dynamically redirects traffic between the Uu and PC5 interfaces. This mechanism aims to mitigate interference between vehicle pairs and VRUs while optimizing the utilization of available resources to meet the QoS requirements of vehicular applications. Algorithm 1 illustrates the logical operation of the Sidelink Manager.

\begin{algorithm}
\caption{Communication Mode Selection}
\begin{algorithmic}[1]
\REQUIRE $\mathcal{P} = \{(V_{i,j}, \omega_i)\}$, $\mathcal{S}$, $\mathcal{W}$

\FORALL{$(V_{i,j}, \omega_i) \in \mathcal{P}$}
    \IF{$v_i \in \mathcal{W}$}
        \IF{$\mathcal{W}[v_i] = \omega_i$}
            \STATE \textit{Remove} $v_i$ \textit{from} $\mathcal{W}$
        \ELSE
            \STATE \textit{Remove corresponding elements from} $\mathcal{P}$ \textit{and} $\mathcal{S}$
        \ENDIF
    \ENDIF
\ENDFOR

\IF{$\mathcal{S} \neq \emptyset$}
    \STATE \textit{Send} $\mathcal{S}$ \textit{to QP xApp and receive predicted modes} $\mathcal{A}$
    \FORALL{$(V_{i,j}, \omega_i), a_i \in \mathcal{P}, \mathcal{A}$}
        \IF{$a_i \neq \omega_i$}
            \STATE \textit{Query resource availability}
            \STATE $R_{\text{UL}}, R_{\text{SL}} \leftarrow \text{\textit{GetResources}}(V_{i,j})$
            \IF{$(a_i = \text{DM} \land R_{\text{SL}} > \theta)$ or $(a_i = \text{IM} \land R_{\text{UL}} > \theta)$}
                \STATE \textit{Apply control:} $\text{\textit{ApplySwitch}}(V_{i,j}, a_i)$
                \STATE \textit{Add} $v_i \rightarrow a_i$ to $\mathcal{W}$
            \ENDIF
        \ENDIF
    \ENDFOR
\ENDIF

\end{algorithmic}
\end{algorithm}

Initially, the manager queries the database and loads the set $\mathcal{P} = \{(V_{i,j}, \omega_i)\}$. $V_{i,j}$ represents the vehicle pairs composed of a source vehicle ($v_i$), a destination vehicle ($v_j$), and $\omega_i$ is the current communication mode (e.g., V2V or V2N), as well as the user and network state denoted by $\mathcal{S}$.

The list $\mathcal{W}$ contains the pairs currently awaiting the application of a control action. The manager removes from $\mathcal{P}$ the pairs whose $v_i$ are already in $\mathcal{W}$ and whose previously recommended actions have not yet been executed, keeping only those eligible for a new decision. If there are remaining valid states $\mathcal{S}$, the manager sends to the QP xApp, which returns a list of actions $\mathcal{A}$ containing the recommended communication modes for each pair $(V_{i,j}, \omega_i)$.

For each recommendation $a_i \in \mathcal{A}$ that differs from the current mode $\omega_i$, the manager verifies the UL and SL resource availability for the pair $V_{i,j}$, obtaining $R_{\text{UL}}$ and $R_{\text{SL}}$. The communication mode switch is applied only if the available resources meet a minimum threshold $\theta$. If this condition holds, the manager executes the action by sending an RIC Control Action to the E2 node, which performs the communication switching via $\text{ApplySwitch}(V_{i,j}, a_i)$. Moreover, it then adds the user to the waiting list $\mathcal{W}$ with the new mode $a_i$. This approach ensures that only feasible decisions are executed, avoiding resource overload and redundant commands for users who have not completed previous transitions.

\section{Proposal Evaluation}
\label{sec:results}

This section evaluates the proposed solution by demonstrating how the strategy mitigates interference, optimizes resource utilization, and minimizes latency and packet loss. For benchmarking, Single-Agent Conservative Q-Learning \textcolor{black}{and two heuristic strategies} are adopted as baselines. \textcolor{black}{The first heuristic baseline is an SINR-based strategy deployed on the xApp, which selects the mode (V2V or V2N) according to a predefined SINR threshold (e.g. 20 dB) and switches to the link with the higher instantaneous SINR, as higher SINR enables more efficient resource usage, as shown in \cite{sinr_based}. The second baseline is the Best CQI strategy deployed by the gNodeB. This strategy selects the mode based on the Adaptive Modulation and Coding (AMC)-estimated bits per resource block, choosing the option with the highest transmission efficiency \cite{best_cqi}.}

\subsection{Evaluation Environment}

We conducted the evaluation using OMNeT++ \cite{omnetpp} with the Simu5G library \cite{simu5g}. The O-RAN support used Near-RT provided by O-RAN Software Community (OSC), more specifically Release I with the OMNeT++ integration proposed in \cite{ho_icoin}. 

\begin{figure}[ht!]
    \centering
    \includegraphics[width=1.0\linewidth]{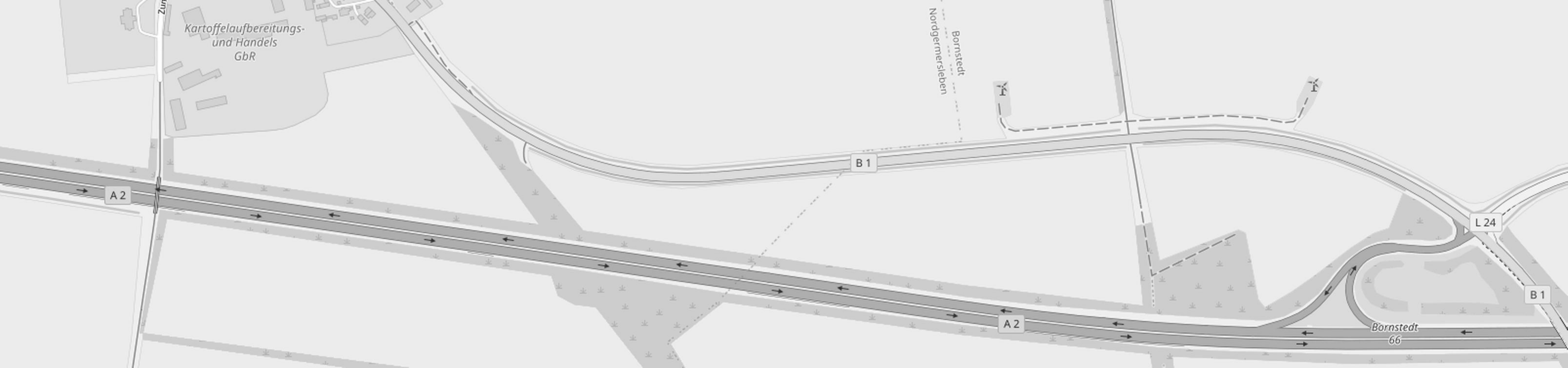}
    \caption{Simulation mobility scenario.}
    \label{fig:sumo-env}
\end{figure}

The simulation scenario is a highway in Germany, as depicted in Figure \ref{fig:sumo-env}. Simulation of Urban MObility (SUMO) \cite{sumo} was used to generate car mobility, simulating 200 vehicles. The network scenario includes a single base station deployed along a 1.5 km segment of the Autobahn highway.

\begin{table}[ht]
    \centering
    \caption{Simulation parameters}
    \begin{tabular}{|l|l|}
    \hline
        \multicolumn{2}{|c|}{\textbf{Network Configuration}} \\
        \hline
            Central Frequency & 6GHz \\
            Number of Cars & 300 \\
            Number of UEs & 10 \\
            OFDM Numerology $\mu$ & 0 \\
            Car/UE Transmit power & 23dBm \\
            Resource Blocks & 50 \\
        \hline
        \multicolumn{2}{|c|}{\textbf{Application Configuration}} \\
        \hline
            Sampling time & 2.5 ms (Constant) \\
            Protocol & User Datagram Protocol (UDP) \\
            Packet size & 1024B \\
            Transmission Time & 20s\\
        \hline
    \end{tabular}
    \label{tab:simulation_parameters}
\end{table}

The scenario implements a see-through application, where each vehicle shares visual information about road conditions with its immediate neighbors. We divided the vehicles into three categories: 150 regular cars (with a maximum speed of 40 m/s), 25 trailers, and 25 coaches (with a maximum speed of 30 m/s for both). The simulation only considers one direction of the highway.

Vehicles transmit 720p and 30fps video streams at approximately 3000 kbps using a constant bit rate (CBR) \cite{video_transmition} application to ensure consistent data flow for the see-through use case while conserving base station resources. In a separate scenario to evaluate pedestrian impact, we introduced 10 static pedestrians, each running a CBR application that simulates music streaming using 80 kbps.

\subsection{Evaluation Methodology}

We structured the evaluation into two phases. The first phase investigates MARL's impact in a pure vehicular IoV scenario without pedestrian presence. The second phase analyzes the coexistence of vehicles and pedestrians, both competing for shared network resources.

We collected the following metrics for both phases:
\begin{itemize}
    \item \textbf{RB Usage Distribution and Interference Mitigation:} Measures how agents balance communication between V2N and V2V links and mitigates the SINR.
    \item \textbf{Latency and Packet Loss:} Captures QoS metrics and quantifies the influence of MARL strategies on performance.
    \item \textbf{Resource Consumption:} Quantifies the computational load associated with model prediction and fine-tuning.
\end{itemize}

\subsection{Use Case 1: Vehicle-Only Scenario}
In this scenario, the goal is to analyze the isolated impact of vehicles with all network resources available for unrestricted use.

\subsubsection{\textcolor{black}{Model Evaluation}}
\begin{figure}[ht!]
    \centering
    \includegraphics[width=1\linewidth]{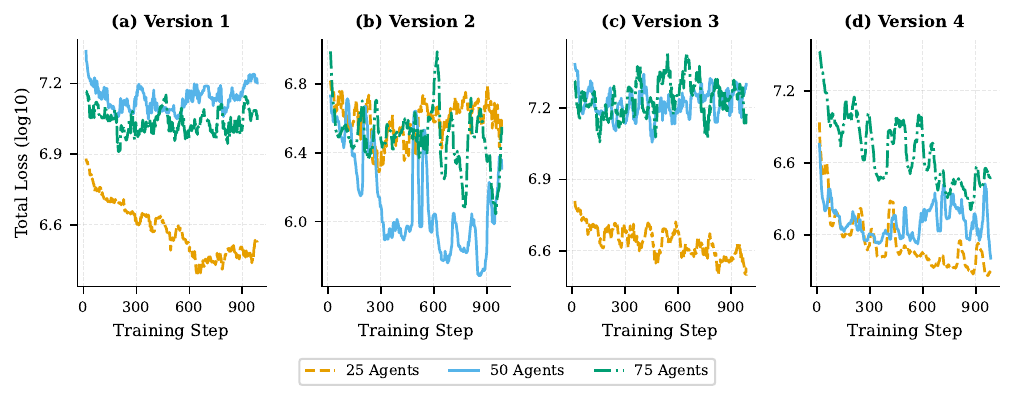}
    \caption{Training loss over versions of the models.}
    \label{fig:loss_iov}
\end{figure}

\textcolor{black}{The evaluation begins by analyzing the agent training process using loss curves across multiple versions. Figure \ref{fig:loss_iov} illustrates this progression and demonstrates that each model produces four versions during network operation. As shown in the figure, the model with 25 agents exhibits the lowest and most stable loss across all versions. This behavior suggests that the policy converged to a more stable region during training, likely because coordinating fewer agents is simpler. Furthermore, a downward trend in loss is observed across versions, indicating a progressive improvement in the policy. Meanwhile, in configurations with 50 and 75 agents, the initial loss exceeds that of the 25-agent model, which is expected given the increased complexity of the joint action space. Across versions, both configurations display alternating periods of instability and convergence, with the 75-agent model exhibiting the most significant fluctuations. However, loss alone does not definitively indicate policy quality. The effectiveness of each model is further assessed during inference, as discussed in the following section.}

\subsubsection{RB Usage Distribution and Interference Mitigation}
In the uplink, the single-agent strategy demonstrated the highest resource usage, exhibiting significant instability as evidenced by the wide interquartile range shown in the boxplot of Figure \ref{fig:enter-label}. This reflects the model's inefficiency in balancing resource allocation. In contrast, the strategy showed considerably lower resource usage for the downlink, as evidenced by the narrow boxplot, suggesting underutilization of infrastructure resources.

On the other hand, multi-agent approaches exhibited more balanced behavior, reducing uplink RB consumption while increasing downlink usage. This indicates a better distribution between V2V (PC5) and V2N (Uu) communications. The reduction in uplink usage suggests that even when using infrastructure resources, the PC5 interface remains constrained by predefined bitmaps, effectively regulating spectrum usage. \textcolor{black}{In contrast, the heuristic baselines preferred V2N usage, leading to higher downlink resource consumption and reduced uplink activity. Such a bias may increase the load on the cellular network and decrease the efficiency of direct V2V link utilization.}

\begin{figure}[ht!]
    \centering
    \includegraphics[width=1\linewidth]{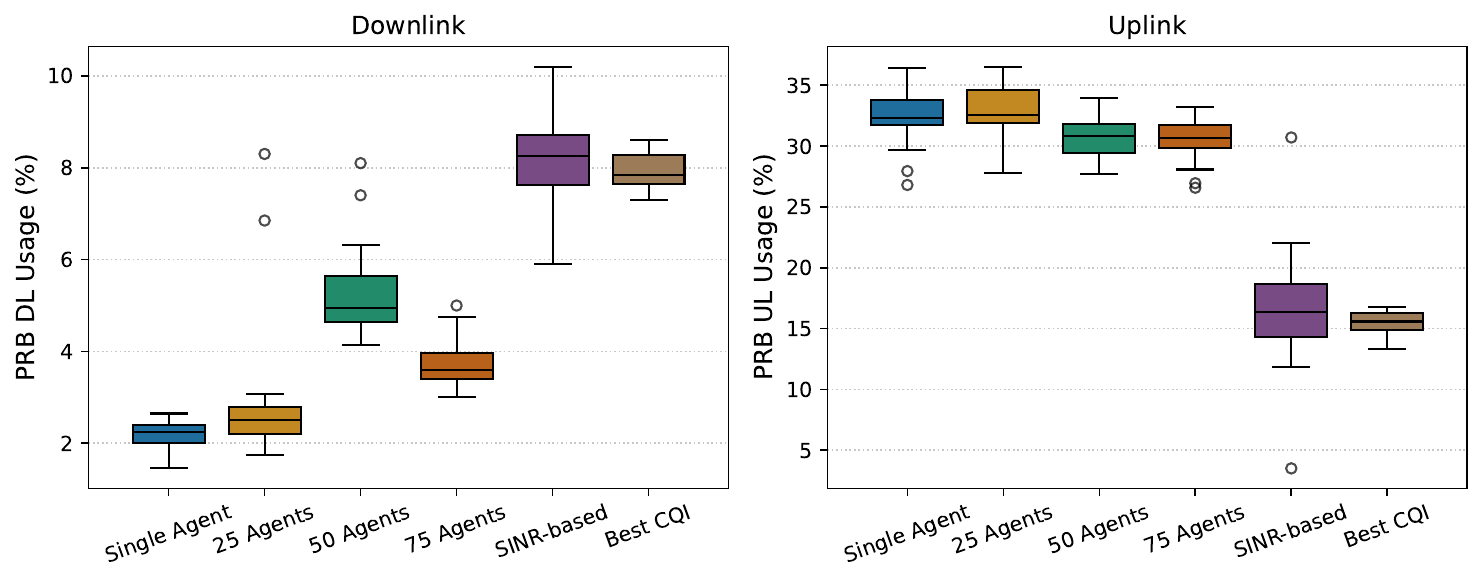}
    \caption{RB usage distribution for uplink and downlink.}
    \label{fig:enter-label}
\end{figure}

\begin{figure}[ht!]
    \centering
    \includegraphics[width=0.65\linewidth]{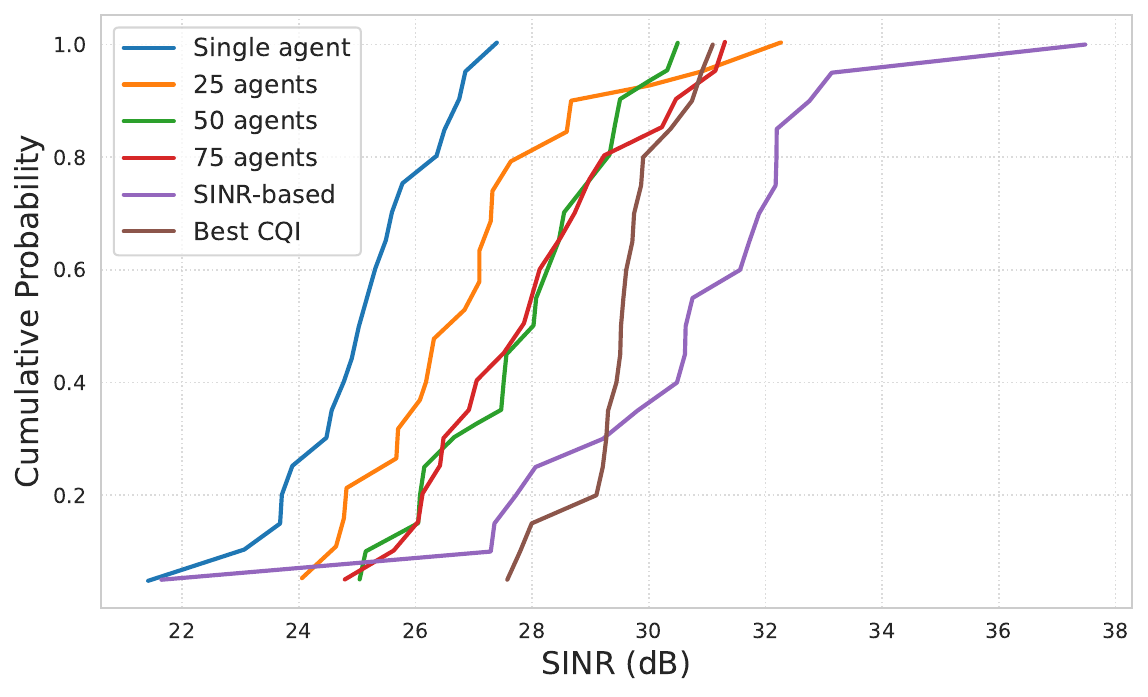}
    \caption{Cumulative distribution function (CDF) of SINR.}
    \label{fig:SINR}
\end{figure}

The importance of balanced communication between V2V and V2N is further highlighted in Figure~\ref{fig:SINR}, which shows improvements in SINR values as the number of agents increases. The single-agent strategy yields the worst SINR results, with only about 30\% of cases achieving an SINR of $\geq$ 25~dB and a maximum probability of approximately 27~dB. In contrast, strategies with more agents, particularly those with 75 agents, show that approximately 80\% of the cases reach a SINR of $\geq$ 25~dB. This demonstrates effective interference mitigation enabled by improved coordination and channel access decisions. \textcolor{black}{Among the heuristic baselines, the SINR-based strategy achieves the highest SINR values, with approximately 80\% of samples at 31.5 dB, as expected, since this method explicitly selects the link with the highest instantaneous SINR. In contrast, the Best CQI strategy demonstrates behavior more similar to the proposed approach with 75 agents, with about 80\% of samples around 30 dB, reflecting its focus on transmission efficiency rather than solely maximizing SINR, which aligns with its greater reliance on V2N communication.}

Among all strategies, the models with 75 and 50 agents stood out for their effectiveness in resource allocation and interference management.

\subsubsection{Latency and Packet Loss}

Regarding QoS metrics, Figure~\ref{fig:enter-label} indicates that the strategy with 50 agents achieves the best performance in terms of both packet loss rate and average latency. The corresponding CDF curves are more left-skewed, indicating a higher probability of achieving lower packet loss and latency values. The 75-agent strategy follows closely, confirming that distributing the decision-making process among agents improves the estimation of Q-values and leads to better action selection for $A_t$.

\begin{figure}[ht!]
    \centering
    \includegraphics[width=1\linewidth]{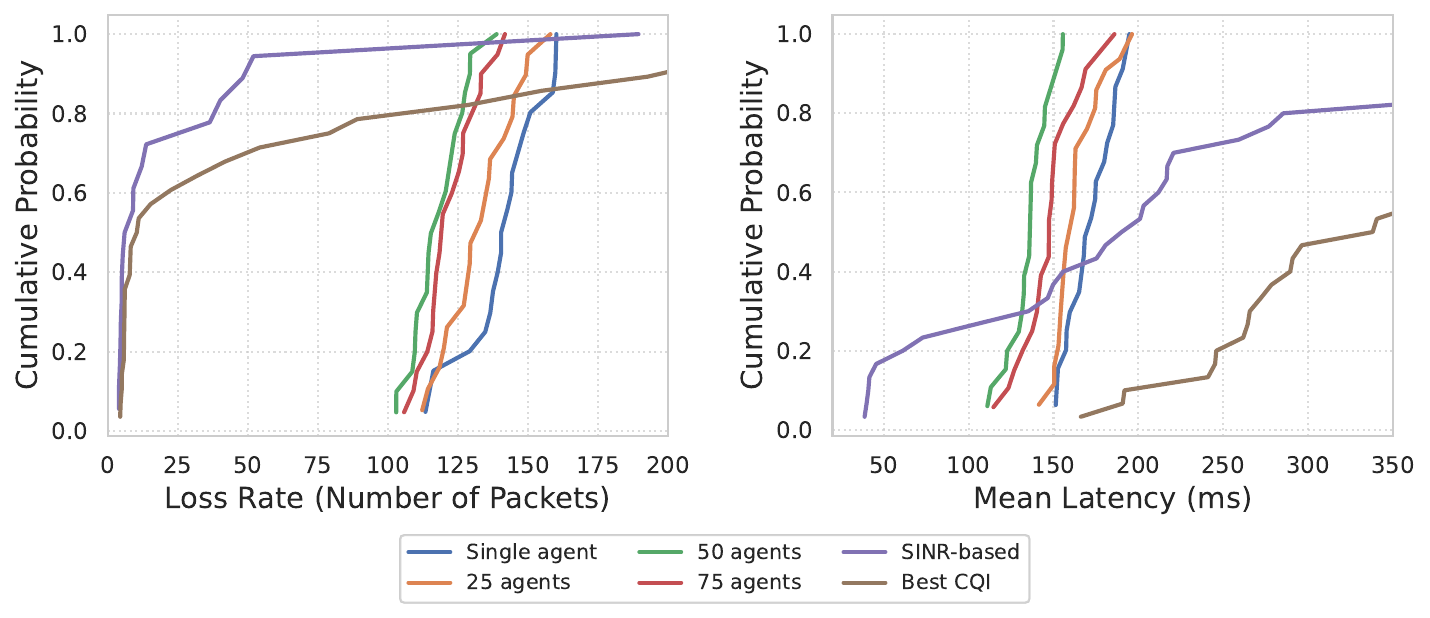}
    \caption{CDF of packet loss and latency.}
    \label{fig:enter-label}
\end{figure}

Moreover, the strategies with fewer agents present higher variability, as seen by the less steep CDF curves, suggesting generalization issues and reduced robustness in decision-making. The heuristic baselines exhibit lower packet loss, particularly for the SINR-based strategy. However, this comes at the cost of significantly higher latency, as the increased reliance on V2N communication introduces additional transmission delays due to the round-trip to the infrastructure. 

\textcolor{black}{When comparing the overall performance of the SINR-based strategy with the 50-agent MARL approach, at the 80th percentile of the cumulative distributions, the SINR-based heuristic improves the SINR by approximately 10.34\%, but at the cost of a latency around 89.65\% higher than that of the 50-agent strategy. These results indicate that maximizing instantaneous SINR does not necessarily lead to better QoS. Instead, the proposed MARL strategy prioritizes lower communication latency by accepting a reduction in SINR, resulting in a better overall balance between radio link quality and application-level performance.} These results reinforce the limitations of single-agent or low-agent-count strategies in complex communication scenarios.

\subsubsection{Computational Cost Analysis}

We compared resource usage during model prediction and fine-tuning to assess the computational costs associated with the decision-making process. The \textit{QP xApp} is responsible for prediction during runtime, while the \textit{ML Management} service handles model fine-tuning. As shown in Table~\ref{tab:cpu_memory_usage}, CPU usage during prediction remains relatively stable regardless of the number of agents, indicating that the inference step is not significantly affected by agent count. However, memory consumption increases with the addition of more agents, reflecting the additional overhead required to manage multiple policies or decision branches.

\begin{table}[ht]
\centering
\caption{Mean Resource Usage During Model Prediction and Update}
\begin{tabular}{@{}lccc@{}}
\toprule
\textbf{Phase} & \textbf{Agents} & \textbf{Memory (MB)} & \textbf{CPU (\%)} \\
\midrule
\multirow{3}{*}{Prediction (QP xApp)} 
    & 25 Agents    &  493.32  & 0.3710 \\
    & 50 Agents    &  691.82  & 0.3977 \\
    & 75 Agents    &  795.29  & 0.3582 \\
\midrule
\multirow{3}{*}{Fine-tuning (ML Mgmt)} 
    & 25 Agents    &  858.85  & 43.83  \\
    & 50 Agents    & 1158.19  & 45.69  \\
    & 75 Agents    & 1456.01  & 44.23 \\
\bottomrule
\end{tabular}
\label{tab:cpu_memory_usage}
\end{table}

In contrast, the fine-tuning phase exhibits a substantial increase in both memory and CPU usage. Still, similar to the prediction step, CPU usage remains relatively comparable across different agent configurations, with memory being the most impacted resource as the number of agents grows. This suggests that while parallel or distributed learning improves performance, it imposes a greater memory burden on the system.

\textcolor{black}{Moreover, during fine-tuning, memory consumption increased by a comparable amount between 25 and 50 agents (about 299 MB) and between 50 and 75 agents (roughly 298 MB), consistent with the linear synchronization cost discussed in Section IV-A. During prediction, memory consumption also increases with the number of agents, but the increment shrinks from nearly 199 MB (25 to 50 agents) to 103 MB (50 to 75 agents), indicating a decreasing growth rate. CPU usage, in turn, remains stable across both phases. This suggests memory usage would continue to scale as the number of agents increases (e.g., 100, 200, and so on).}

To evaluate the trade-off between QoS improvements and computational resource costs and to ensure fair comparisons among metrics with different units and scales, such as packet loss, delay, CPU usage, and memory consumption, the analysis normalizes all values to the range $[0, 1]$ using min-max normalization:

\begin{equation}
X_{\text{norm}} = \frac{X - X_{\min}}{X_{\max} - X_{\min}}
\end{equation}

After normalizing the values $\left(X_{\text{norm}}\right)$, the analysis computes two key indicators: QoS efficiency and computational cost. For QoS efficiency, the process uses the normalized values of packet loss $\left(X_{\text{norm\_loss}}\right)$ and latency $\left(X_{\text{norm\_latency}}\right)$. Since lower values indicate better performance in these metrics, the analysis inverts the scale after normalization to maintain consistency. It calculates the mean of the normalized values and applies the transformation $1 - \text{mean}\left(X_{\text{norm}}\right)$, which keeps the result within the $[0, 1]$ range and ensures that higher values represent better QoS performance.

To assess computational cost, the analysis calculates the average of the normalized CPU usage $\left(X_{\text{norm\_cpu}}\right)$ and memory usage $\left(X_{\text{norm\_memory}}\right)$. In this case, higher values correspond to higher resource consumption and, therefore, greater computational cost.

\begin{figure}[h!]
    \centering
    \includegraphics[width=0.7\linewidth]{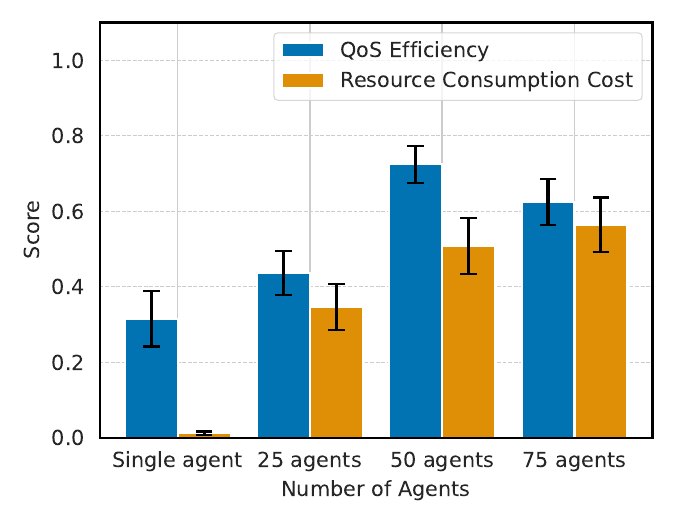}
    \caption{Trade-off between QoS efficiency and computational cost on IoV only scenario.}
    \label{fig:iov_score}
\end{figure}

Figure \ref{fig:iov_score} presents the results for the analyzed indicators. Regarding the QoS efficiency indicator, the configuration with 50 agents achieved the best performance, reaching a score of approximately 0.726. In comparison, the single-agent and 25-agent configurations exhibited reductions of 56.6\% and 39.8\%, respectively.

Increasing the number of agents to 75 reduced efficiency by around 14\% compared to the peak value achieved with 50 agents, suggesting potential overhead or decreased coordination among agents.

When comparing the configurations that delivered the best results in each indicator, 50 agents for QoS efficiency and a single agent for computational cost, the single-agent setup consumed approximately 97.7\% fewer resources. This finding highlights the significant computational savings of the single-agent approach despite its notable loss in QoS efficiency.

Regarding the cost of resource consumption, the 75-agent configuration incurred the highest overhead. In contrast, the configurations with 50 agents, 25 agents, and a single agent consumed approximately 10.3\%, 38.8\%, and 97.9\% fewer resources, respectively. These results show that increasing the number of agents improves QoS efficiency up to a certain point but also leads to a substantial rise in resource usage.

\subsection{Use Case 2: Vehicular-Pedestrian Coexistence in IoV}

In this scenario, we assess the impact of coexistence between pedestrians and vehicles in IoV environments, where both compete for network resources.

\subsubsection{\textcolor{black}{Model Evaluation}}

\begin{figure}[ht!]
    \centering
    \includegraphics[width=1\linewidth]{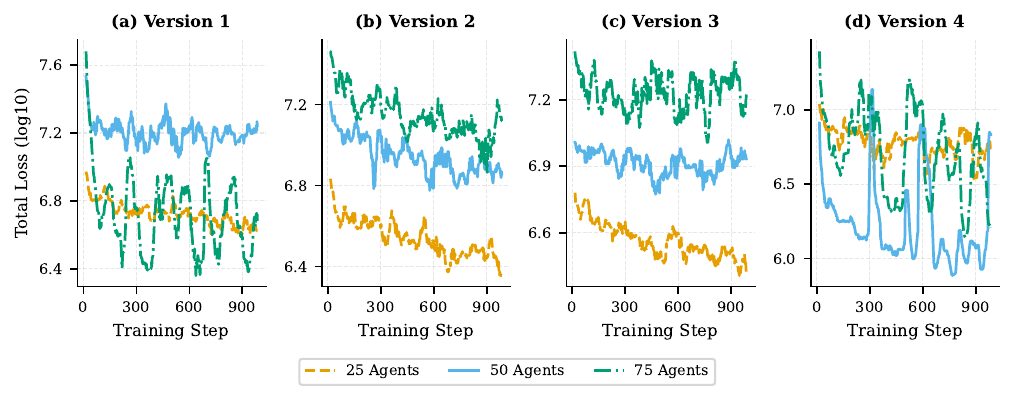}
    \caption{Training loss over models versions.}
    \label{fig:loss_ped}
\end{figure}

\textcolor{black}{Figure \ref{fig:loss_ped} shows the training loss for Vehicular-Pedestrian Coexistence on the same infrastructure scenario. Compared to the scenario with only IoV, the introduction of pedestrians adds a new level of complexity to the coordination task, reflected in the loss dynamics across all agent configurations and versions.}

\textcolor{black}{Unlike the previous scenario, all three configurations exhibit greater variability in losses across versions. In the first model version, the 75-agent model stands out, showing lower losses than the 50-agent model, but with high instability. This suggests that the presence of pedestrians alters the previously observed initial convergence behavior, leading the policy to explore a wider region of the action-state space before stabilizing.}

\textcolor{black}{In versions 2 and 3, the models show signs of stabilization, with the 25-agent configuration again showing the most consistent convergence. The 50 and 75-agent models continue to exhibit oscillatory behavior in later versions, suggesting greater difficulty in converging in an environment with more users competing for resources. As in the previous scenario, these loss curves characterize training dynamics but do not directly quantify policy quality.}

\subsubsection{RB Usage Distribution and Interference Mitigation}

\begin{figure}[ht!]
    \centering
    \includegraphics[width=1\linewidth]{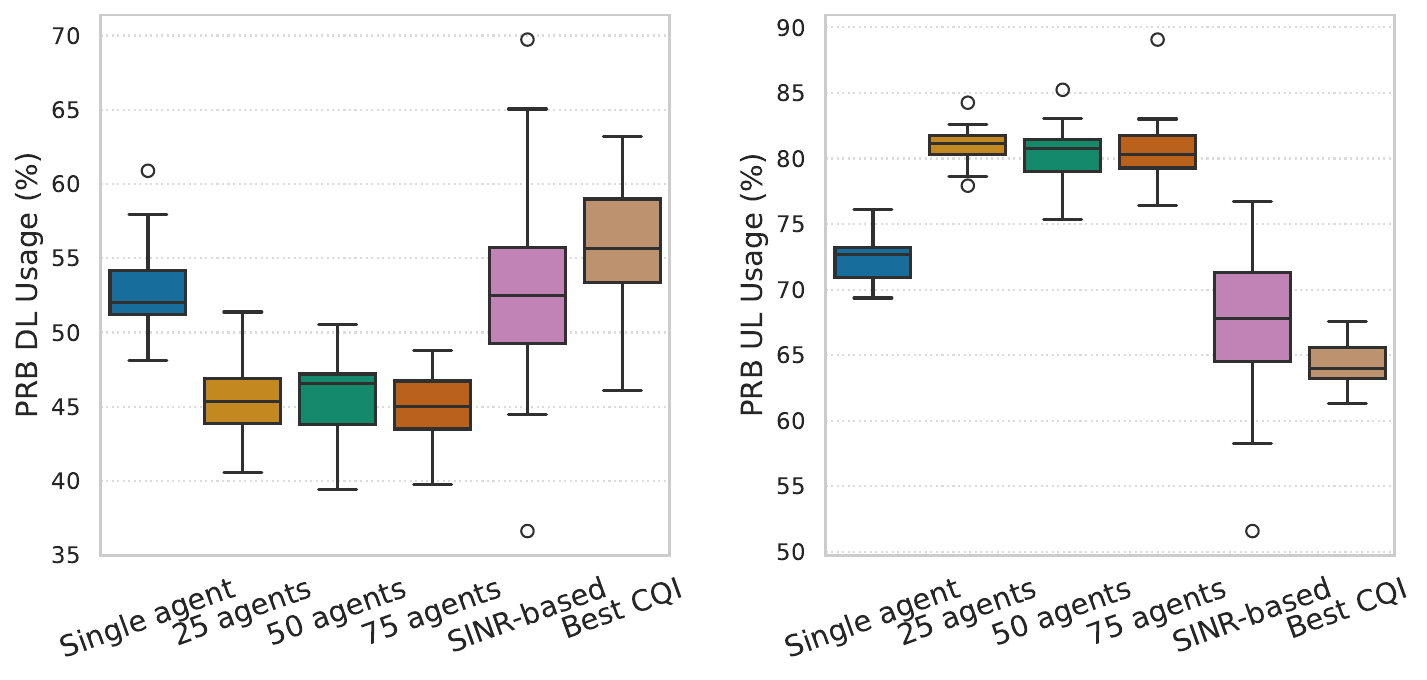}
    \caption{RB usage distribution in vehicular-pedestrian coexistence.}
    \label{fig:usage_pedestre}
\end{figure}

Figure \ref{fig:usage_pedestre} illustrates the RB usage distribution in the presence of both vehicles and pedestrians. Compared to the previous vehicular-only scenario, the overall resource consumption increased, particularly in the downlink, which is consistent with the additional communication demands introduced by pedestrian entities. However, the multi-agent strategies continue to show improved stability in resource allocation, with narrower boxplots and more balanced uplink and downlink utilization. 

\begin{figure}[ht!]
    \centering
    \includegraphics[width=0.65\linewidth]{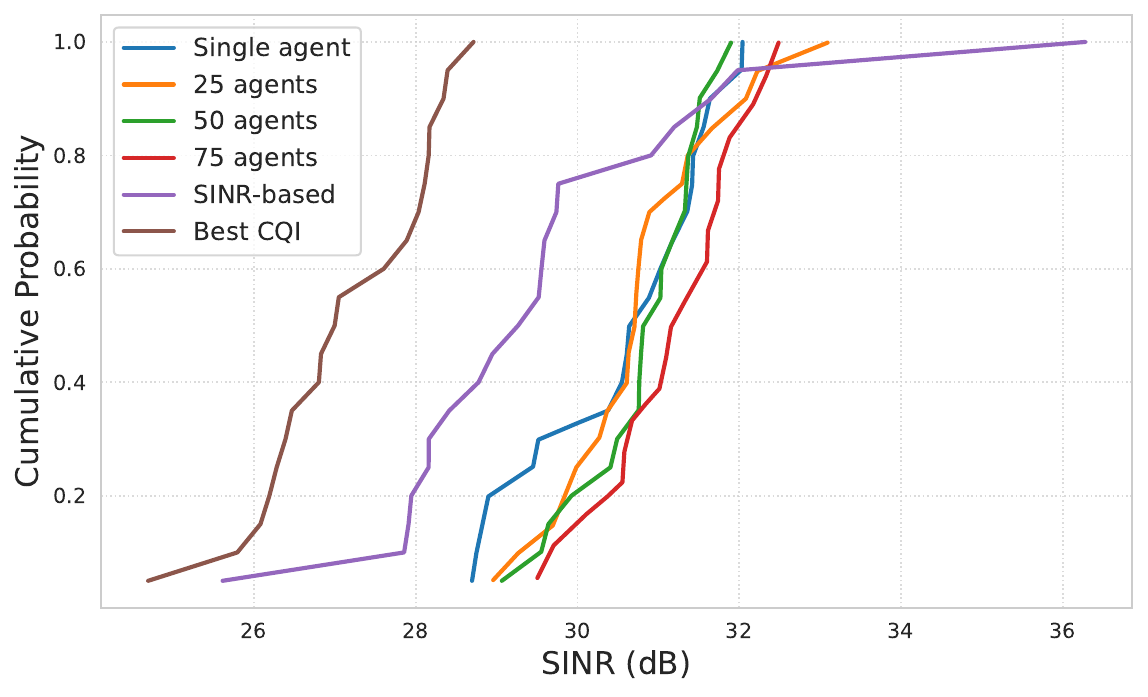}
    \caption{CDF of SINR in vehicular-pedestrian coexistence.}
    \label{fig:SINR_ped}
\end{figure}

Figure~\ref{fig:SINR_ped} presents the SINR distribution for this scenario. The single-agent configuration remains the least effective, with most samples having a level below 31 dB between the RL models. On the other hand, the 75-agent strategy significantly outperforms the others, with a large portion of samples above 32~dB, indicating successful interference mitigation through distributed coordination. The curves suggest that increasing the number of agents helps better manage interference in high-density environments with heterogeneous actors.

\textcolor{black}{Nevertheless, the heuristic baselines exhibit the lowest SINR performance in this scenario. This behavior is mainly due to increased infrastructure load from pedestrian users, which intensifies interference levels. As a result, simple mode-selection strategies based solely on a single metric (e.g., SINR or CQI) are insufficient to properly balance resource usage and interference, highlighting the importance of more coordinated, adaptive approaches.}

\subsubsection{Latency and Packet Loss}

\begin{figure}[ht!]
    \centering
    \includegraphics[width=1\linewidth]{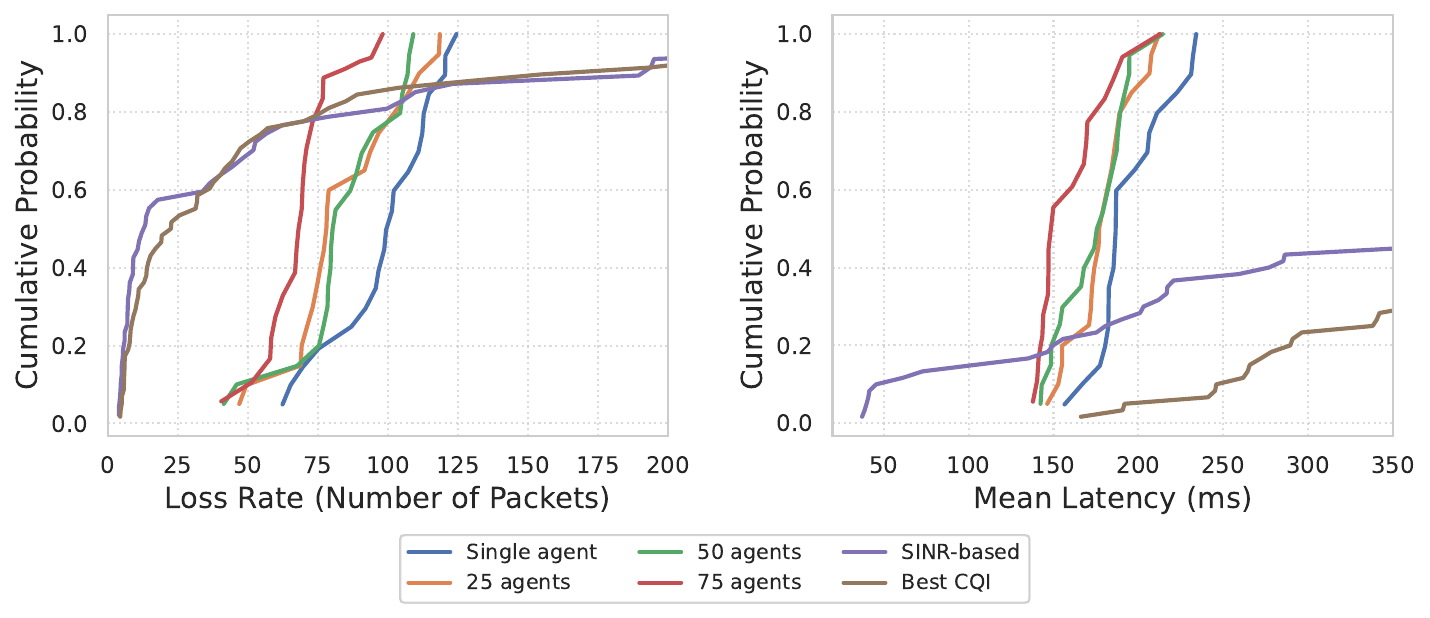}
    \caption{CDF of packet loss and latency in vehicular-pedestrian coexistence.}
    \label{fig:qos_ped}
\end{figure}

The QoS analysis for the coexistence scenario is shown in Figure~\ref{fig:qos_ped}. Similar to the SINR results, the 50 and 75-agent configurations provide the best performance, with their CDF curves for both packet loss rate and mean latency shifted further to the left. This indicates a higher probability of achieving lower latency and fewer lost packets.

\textcolor{black}{On the other hand, SINR-based and Best CQI present the worst overall QoS performance. As shown in the right plot, neither curve reaches a cumulative probability above 0.45 within the evaluated latency range, indicating that a significant portion of transmissions experience excessively high mean latency, exceeding 200 ms in most cases. Similarly, in the loss rate plot, both baselines accumulate probability mass at lower values early on but plateau well below 1.0, suggesting a long tail of high-loss events.} This result reinforces the conclusion that increasing the number of agents improves communication efficiency and overall reliability, especially in more complex and competitive environments such as those involving vehicles and pedestrians.

\subsubsection{Computational Cost Analysis}

Table~\ref{tab:cpu_mem_pedestrian} summarizes the average memory and CPU consumption during the two main operational phases: model prediction and fine-tuning. 

\begin{table}[ht]
\centering
\caption{Mean Resource Usage During Model Prediction and Update}
\begin{tabular}{@{}lccc@{}}
\toprule
\textbf{Phase} & \textbf{Agents} & \textbf{Memory (MB)} & \textbf{CPU (\%)} \\
\midrule
\multirow{3}{*}{Prediction (QP xApp)} 
    & 25 Agents   & 470.86 & 0.90 \\
    & 50 Agents   & 685.15 & 0.39 \\
    & 75 Agents   & 734.46 & 0.48 \\
\midrule
\multirow{3}{*}{Fine-tuning (ML Mgmt)} 
    & 25 Agents   & 884.14 & 48.85 \\
    & 50 Agents   & 1211.24 & 55.10 \\
    & 75 Agents   & 1577.82 & 50.67 \\
\bottomrule
\end{tabular}
\label{tab:cpu_mem_pedestrian}
\end{table}

During prediction, memory consumption increases with the number of agents, reaching a maximum of 734MB for 75 agents. CPU usage remains relatively low across all configurations, with the 25-agent setup consuming the most CPU, approximately 0.90\%. This indicates that increasing the number of agents imposes a greater memory burden during runtime inference while maintaining modest CPU demands.

In contrast, during fine-tuning, memory and CPU usage increase significantly with the number of agents. The 50-agent configuration consumed the highest average CPU at 55.10\%, followed by the 75-agent configuration at  50.67\% and the 25-agent configuration at 48.85\%. Regarding memory usage, it also increased proportionally, from 884MB in 25 agents to 1578MB with 75 agents, reflecting the cost of maintaining multiple parallel learning processes.

These findings highlight that the fine-tuning phase is considerably more resource-intensive than prediction, particularly in terms of CPU usage. While multi-agent learning enhances decision-making performance, it incurs a higher computational demand during the training phase. \textcolor{black}{Furthermore, memory consumption increased by approximately 327 MB between 25 and 50 agents and 367 MB between 50 and 75 agents, reflecting a slightly higher growth rate than that observed in the vehicle-only scenario. During prediction, memory consumption also increases with the number of agents, but the increment shrinks from nearly 214 MB to around 49 MB, indicating a decreasing growth rate. CPU usage, in turn, does not follow a monotonic trend in either phase, unlike the vehicle-only scenario. This suggests memory usage would continue to scale as the number of agents increases (e.g., 100, 200, and so on).}

\begin{figure}[h!]
    \centering
    \includegraphics[width=0.7\linewidth]{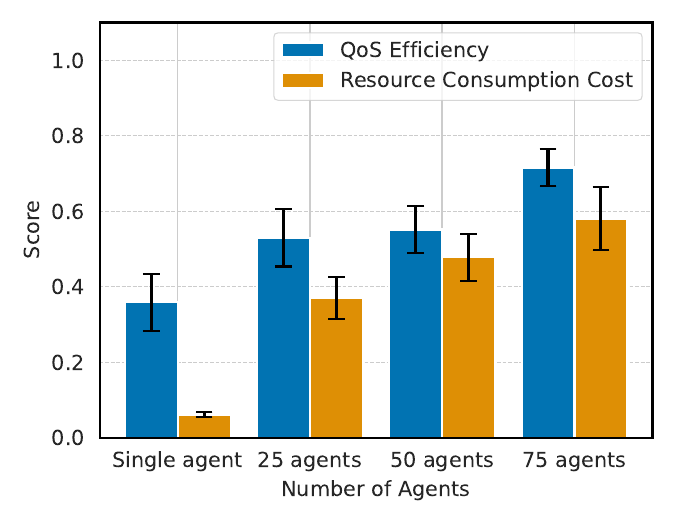}
    \caption{Trade-off between QoS efficiency and computational cost in vehicular-pedestrian coexistence.}
    \label{fig:score_ped}
\end{figure}

Finally, Figure \ref{fig:score_ped} presents the obtained results of the trade-off between QoS efficiency and computational cost in a vehicular-pedestrian coexistence scenario. Regarding the QoS efficiency indicator, the configuration with 75 agents achieved the best performance, reaching a score of approximately 0.715. In comparison, the 50-agent, 25-agent, and single-agent configurations exhibited reductions of 22.94\%, 26.26\%, and 49.82\%, respectively. In this use case, the results show that increasing the number of agents improves QoS efficiency, as this scenario is more complex due to heterogeneity than use case 1, which is more homogeneous. 

Concerning the cost of resource consumption, the 75-agent configuration also incurred the highest overhead. In comparison, the configurations with 50 agents, 25 agents, and a single agent consumed approximately 17.76\%, 36.56\%, and 89.52\% fewer resources than the  75-agent configuration, respectively. These results reinforce that increasing the number of agents improves QoS efficiency but also leads to a substantial rise in resource usage. Therefore, achieving an optimal trade-off between performance and cost requires careful tuning.

\textcolor{black}{\section{Open Challenges and Future Research Directions}}

\textcolor{black}{This section discusses open challenges and future research directions that build upon the proposed framework. While the presented approach provides a basis for intelligent V2X mode selection within the O-RAN paradigm, several aspects related to real-world deployment, dynamic network conditions, and large-scale operation remain to be further investigated. Addressing these aspects represents a natural continuation of this work and may enhance its applicability in other scenarios.}

\textcolor{black}{\subsection{Real-World Experimental Validation}}

 \textcolor{black}{The present proposal leverages real-world containerized intelligent controllers from the OSC. This approach facilitates integration into experimental environments compatible with the O-RAN architecture. Although the proposed solution was validated through simulation-based experiments that demonstrated its effectiveness, real-world validation poses significant challenges. In this context, testbeds based on software-defined radios (SDRs) and open-source platforms provide promising alternatives. Nevertheless, existing open-source RAN platforms offer limited sidelink support, and the availability of compatible commercial devices is scarce, which restricts large-scale experimentation. Additionally, experiments in real IoV settings are constrained by safety considerations and deployment complexity. These factors underscore the necessity for hybrid validation methodologies that integrate simulation with real-world experimentation.}

\textcolor{black}{\subsection{Channel Aging in Control Loops}}

 \textcolor{black}{In high-mobility scenarios, rapid channel variations and frequent handovers can introduce significant channel aging. Consequently, the state observed by the Near-RT RIC may be outdated by the time a control action is applied, resulting in suboptimal decision-making. Since these control loops operate on timescales ranging from 10 ms to 1 s, this latency can result in a stale-state problem. Understanding these time constraints in high-mobility environments remains an open challenge. Future research should investigate these degradation scenarios and explore mitigation strategies, such as predictive state estimation and aging-aware reward shaping, to enhance the proposed solution.}

\textcolor{black}{\subsection{Out-of-Coverage Operation}}

\textcolor{black}{The proposed framework considered continuous network coverage for UE measurement reporting and RIC control actions. However, out-of-coverage and intermittent connectivity scenarios are critical for V2X safety applications and remain unresolved challenges. These scenarios may benefit from hybrid approaches, in which the RIC optimizes long-term policy while UEs retain autonomous decision-making for immediate safety in the absence of network connectivity. Future research should examine coordination between these layers, emphasizing policy consistency and system stability during transitions between centralized and distributed control modes.}

\textcolor{black}{\subsection{Multi-Cell Coordination and Interference Management}}

Our study focuses on a single-cell environment. Future work consists of extending the framework to multi-cell deployments, which introduces significant challenges in coordination and interference management. Sidelink transmissions may affect users in neighboring cells, especially with aggressive resource reuse. While recent 3GPP releases improved sidelink communication, efficient inter-cell coordination and group-based resource management remain active research areas. This highlights the need to consider multi-cell O-RAN deployments and their interactions with handover mechanisms under dynamic, heterogeneous network conditions. \textcolor{black}{Furthermore, future evaluations should include a larger number of agents to assess the performance of the proposed solution in more complex, large-scale scenarios.}

\section{Conclusion}
\label{sec:conc}

This work presented a MARL-based solution for determining the communication mode between PC5 interfaces, using either Sidelink (V2V) or network-based communication (V2N) via the Uu interface. The objective of the proposed solution was to select the optimal communication mode to mitigate interference among peers and improve the utilization of available radio resources. We tested two scenarios: the first involving only vehicles utilizing all available resources, and the second considering coexistence with VRUs. The results indicate that MARL-based solutions effectively mitigate interference in both cases, optimize resource usage, and meet the QoS metric requirements.

\section*{Acknowledgment}
This work was supported by the National Council for Scientific and Technological Development (CNPq) - Research Productivity Fellowship (Grant No. 313083/2023-1) and Pernambuco Research Foundation (FACEPE) (Grant No. IBPG-0130-1.03/23).

{\appendix[Model Descriptor]
\begin{lstlisting}[language=JSON]
{
    "name": "cqmix",
    "version": "1.0",
    "runtime": {
        "name": "python",
        "encoding": "protobuf",
        "version": "3.10.12",
        "dependencies": {
            "Pip": {
                "requirements": [
                    {
                        "name": "torch",
                        "version": "2.5.1"
                    },
                    {
                        "name": "pandas",
                        "version": "2.2.2"
                    },
                    {
                        "name": "numpy",
                        "version": "1.26.4"
                    }
                ]
            }
        }
    },
    "methods": [
        {
            "choose_com_mode": {
                "input": "NetworkState",
                "output": "CommunicationModeType",
                "description": "Receives network and user status and sets communication mode"
            }
        }
    ]
}
\end{lstlisting}}

\vspace{11pt}
\begin{IEEEbiography}[{\includegraphics[width=1in,height=1.25in,clip,keepaspectratio]{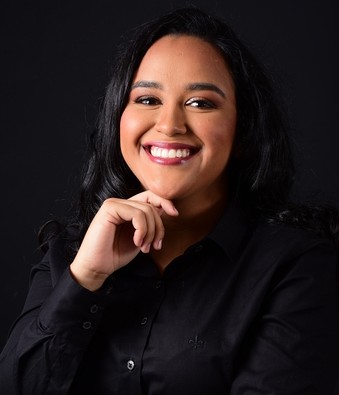}}]{Maria Katarine Santana Barbosa}
received the B.Sc. degree in Electrical Engineering from University of Pernambuco (UPE), Recife-Brazil, in 2021. She received M.Sc in Computer Science from the Informatics Center of the Federal University of Pernambuco (CIn/UFPE), in 2023.  She is currently pursuing the PhD degree in Computer Science in Universidade Federal de Pernambuco. Her current research interests are Optimization, Network Functions Virtualization, Software-defined Radio, 5G and 6G networks.
\end{IEEEbiography}

\vspace{11pt}

\begin{IEEEbiography}[{\includegraphics[width=1in,height=1.5in,clip,keepaspectratio]{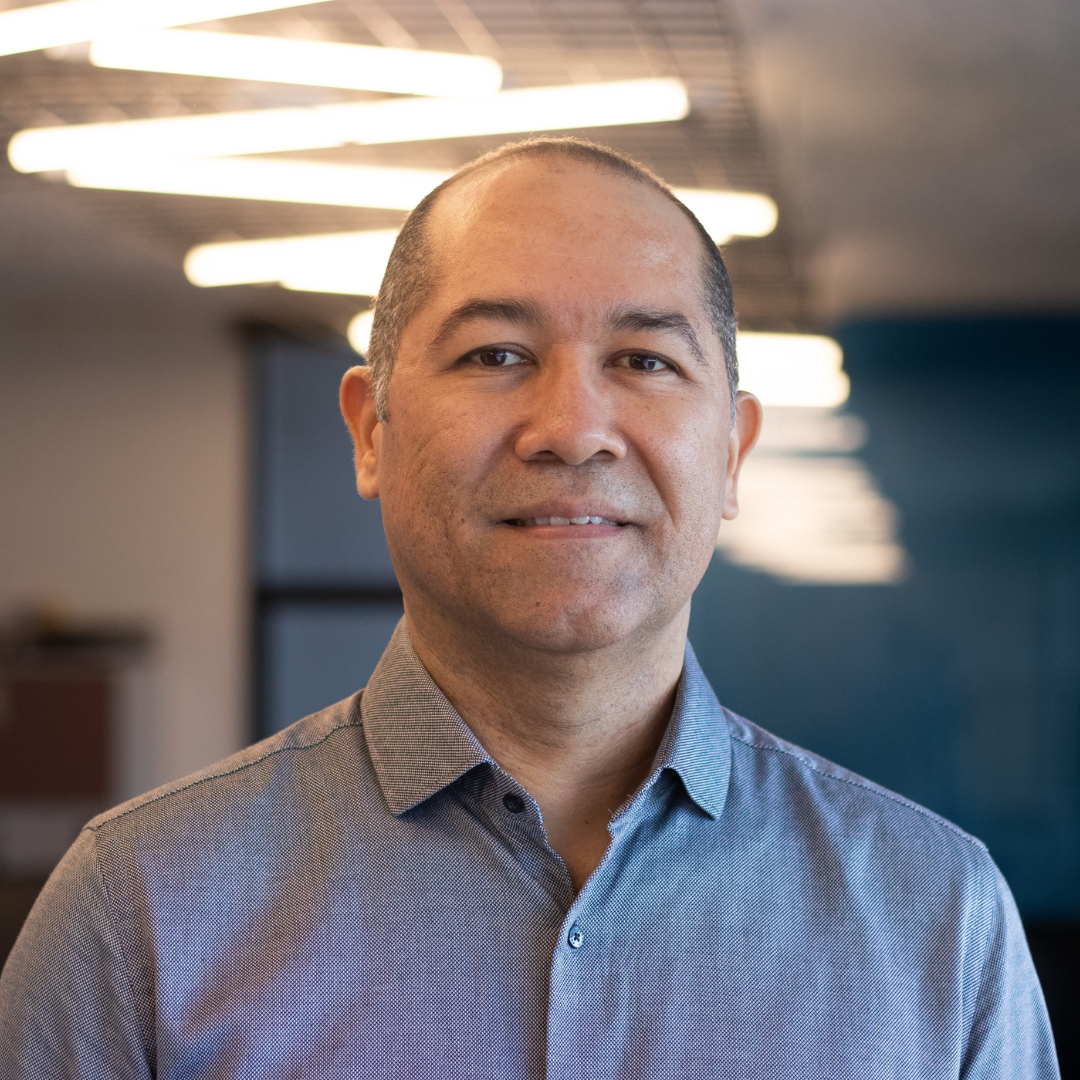}}]{Kelvin Lopes Dias}
received a Ph.D. degree in Computer Science from Federal University of Pernambuco (UFPE), Recife-Brazil, in 2004. He is a full professor at the Center of Informatics (CIn) of the UFPE. His current research interests are Software-defined Networking, Network Functions Virtualization, Edge Computing, 6G networks, and Advanced \& Intelligent Network Architectures.
\end{IEEEbiography}

\vfill

\end{document}